\documentclass[]{revtex4}  
\usepackage{times}
\usepackage{graphicx}
\usepackage{times}

\begin{document}
\title{Inferring the diameter of a biopolymer from its stretching response}

\author{Ngo Minh Toan$^{1,2}$, Davide Marenduzzo$^{3,4}$ and
Cristian Micheletti$^1$\\
\small$^1$ International School for Advanced Studies (S.I.S.S.A.) and
INFM, Via Beirut 2-4, 34014 Trieste, Italy\\
\small$^2$ Institute of Physics, 10 Dao Tan, Hanoi, Vietnam\\
\small $^3$ the Rudolf Peierls Centre for Theoretical Physics, Oxford University, 1 Keble Road, Oxford OX1 3NP, United Kingdom\\
\small $^4$ Mathematics Institute, University of Warwick, Coventry CV4 7AL,
United Kingdom}

\begin{abstract}
We investigate the stretching response of a thick polymer model by
means of extensive stochastic simulations. The computational results
are synthesized in an analytic expression that characterizes how the
force versus elongation curve depends on the polymer structural
parameters: its thickness and granularity (spacing of the monomers).
The expression is used to analyze experimental data for the
stretching of various different types of biopolymers: polypeptides,
polysaccharides and nucleic acids. Besides recovering elastic
parameters (such as the persistence length) that are consistent with
those obtained from standard entropic models, the approach allows to
extract viable estimates for the polymers diameter and
granularity. This shows that the basic structural polymer features
have such a profound impact on the elastic behaviour that they can
be recovered with the sole input of stretching measurements.
\end{abstract}
 
\maketitle

\section{Introduction}
In the past decade a remarkable technological progress has allowed
experiments where single polymers are manipulated by means of optical
or magnetic tweezers, micro-needles or atomic force microscopes
 \cite{bustamante03,busta,single2,strick}. These micro-manipulation
techniques were used to characterize with great accuracy how various
biopolymers such as DNA, proteins and polysaccharides elongate under
the action of a stretching force
 \cite{block1,williams,strick,linke,linke2,fernan,gaub99,marsza2,clarke1,clarke2,wang}. The
wealth of collected experimental data have constituted and still
represent an invaluable and challenging benchmark for theoretical
approaches to the problem
 \cite{Boal,pincus,WLC2,cieplak10,cieplak13,schulten1,schulten3}. The typical
models introduced to account for the observed stretching measurements
rely on a description of the polymer as a one-dimensional curve with a
certain persistence length. Two celebrated examples are provided by
the freely-jointed and the worm-like chain (FJC and WLC,
respectively). Their widespread use reflects both their success in
fitting experimental measurements and their simplicity of
formulation. In fact, within the FJC model the polymer is described as
a succession of phantom independent segments \cite{flory,Boal} while
the WLC relies on a description of the chain as a continuous curve
characterized by a local bending rigidity
 \cite{WLC1,WLC2,podg01}.

In this study we introduce a mesoscopic approach where a polymer is
specified by its basic three-dimensional structural properties which,
in turn, dictate how the chain responds under mechanical tension. This
provides a novel view on the problem where the persistence length of
the polymer (and its entire stretching curve) is not determined
phenomenologically but is specified by fundamental structural
parameters. As we shall discuss, this perspective opens the
possibility to extract the fundamental structural parameters of
biopolymers given the measurements of their stretching response.

The feasibility of the scheme outlined above relies on the possibility
to model in an effective way how the polymer structural features
impact on the elastic response. As a pre-requisite, any such model
should go beyond the one-dimensional schematization of the chain and
capture its three-dimensional steric hindrance and the granularity
associated to the monomeric units. Arguably, the simplest theoretical
framework that possesses these requirements is the thick-chain (TC)
model that has been recently developed in geometrical contexts
motivated by the knotting of circular DNA
 \cite{buck1,rawd99,Thickness1,Thickness2,Thickness3,opthelix}. In the
TC framework a polymer is modelled as a discretised tube with a given
thickness. The steric effects associated to the finite thickness are
fully taken into account and introduce some strict constraints for the
viable configurations that the chain can attain in the
three-dimensional space. The implications of the selection in
structure space operated by this steric criterion have been
investigated in a variety of biophysical problems. In particular it
has allowed to elucidate the ubiquitous emergence of secondary motifs
in proteins and the characterization of DNA knotting probabilities and
DNA packaging thermodynamics \cite{opthelix,dnapack,cozzarelli}. The
diversity of these biological problems suggests that the thick chain
model may be a sufficiently general framework to capture the
stretching response of biopolymers with very different thickness and
persistence length such as DNA, proteins and polysaccharides.

A detailed discussion of the model is provided in the next section
where we also discuss how its stretching response was characterized by
computational means leading to a parametrization of the stretching
curve in terms of the chain contour length, its thickness and the
length of the modular units (monomers). The resulting parametric
expression is finally used to fit the experimental data on three
different instances of biopolymers: the PEVK-domain of protein
 \cite{linke,linke2,fernan}, cellulose \cite{marsza1,marsza2,marsza3}
and double-stranded DNA \cite{block1}. It is shown that this approach
allows to extract meaningful quantitative estimates of the effective
thickness and modular granularity of the polymers. The persistence
length implied by the structural parameters is consistent with that
obtainable from the WLC or, according to the contexts, the FJC. In
addition the overall quality of the fit of experimental data (measured
in terms of $\chi^2$) improves that achievable with either of the two
reference entropic models.

\section{Theory and Methods}

To characterize the impact of a polymer's finite thickness on the
stretching response we shall view the polymer as a tube with a uniform
circular section. Consistently with other studies we define the chain
thickness, $\Delta$, as the radius of the circular section and,
furthermore, we coarse-grain the polymeric chain into a succession of
discrete units equispaced at distance $a$. From a schematic point of
view the discretised thick chain can be thought of as consisting of
tethered discs or oblate ellipsoids of radius $\Delta$ and spacing
$a$, as in Fig. 1a.

Within the framework of elasticity theory, the elastic properties of a
homogeneous tube are well-characterised in the case where the tube is bent
with respect to the straight reference configuration. If the bending
angle is small then the leading term in the elastic energy penalty has
a quadratic dependence on the bending curvature. Already at this
perturbative stage the thickness of a polymer, viewed as a
uniformly-dense tube, controls the elastic response since it
determines the bending rigidity coefficient, $\kappa_b$. In fact,
throughout a vast class of biopolymers, $\kappa_b$ shows an
approximate quartic dependence on $\Delta$ which implies that thermal
excitations will be unable to excite appreciable local bends in
sufficiently thick polymers (as microtubules) \cite{Boal}. By
converse, for smaller thickness, thermal excitations may cause
appreciable deviations from the straight configuration. In this case
the quadratic expansion of the elastic energy may be questioned since
it does not penalize enough highly-bent conformations. In the case of
peptides, for example, the persistence length at room temperature is
comparable with the nominal
thickness \cite{gaub97,gaub99,opthelix}. From the quadratic
approximation of the elastic energy one would conclude that thermal
excitations should easily introduce deformations corresponding to
bending radii smaller than $\Delta$ -- a fact that should be strictly
forbidden since it would lead to self-intersections. It therefore
appears that a general mesoscopic model for the elasticity of
biopolymers should take into account explicitly how the accessible
conformational space is limited by the polymer finite thickness.

A theoretically appealing and computationally effective framework for
treating the finite thickness in discrete polymer models was recently
proposed in the context of ideal shapes of knots
 \cite{buck1,buck2,Thickness2,rawd99,Thickness1,raw2000,Thickness3}.

{Physically, the finite thickness introduces
restrictions on polymer chains that are both of local and non-local
character. To avoid local self-intersection of the thick tube, the
local radius of curvature must be no smaller than $\Delta$
 \cite{buck1,Thickness1}, as anticipated above. This introduces a
limitation for the range of attainable values for the angle formed by
two consecutive bonds of the discretised thick chain. Besides this
local effect there is also a non-local one: any two portions of the
tube, at a finite arc-length separation, cannot interpenetrate
 \cite{buck1,Thickness1}.} In traditional beads-and-strings models it
is only this second effect that is taken into account through a
pairwise potential with a hard-core repulsion. Interestingly, one
needs to go beyond pairwise interactions to account for the above
mentioned effects in discretized polymer chains \cite{secstr,JPC}. In
fact, the requirement on the local radius of curvature can be enforced
by finding the radii of the circles going through any consecutive
triplet of points and ensuring that each of them is greater than
$\Delta$. The non-local effect can be addressed within the same
framework by requiring that the minimum radius among circles going
through any non-consecutive triplet of points is also greater than
$\Delta$. Indeed, the finite thickness, $\Delta$, of the discretized
tube constrains the radii of the circles going through any triplet of
distinct points to be greater than $\Delta$ (see Fig. 1b) 
 \cite{Thickness1}.

The treatment followed here therefore focuses on thickness-induced
effects that are complementary to those captured by the WLC. In fact,
in the latter approach all three-dimensional arrangements of a chain
are permitted but each structure carries a statistical weight that
reflects its degree of local bending. In the thick chain model the
chain configurations that are incompatible with the finite polymer
thickness are excluded but allowed configurations have the same
statistical weight. The parameters that characterise a thick discrete
chain within this model are three: the chain contour length, $L_c$,
its thickness, $\Delta$, and the spacing, $a$, of the monomers which
control the chain discretisation (granularity). It may be envisaged
that a more powerful scheme would be to combine the WLC and TC by
introducing a bending rigidity penalty among the physically allowed
structures. This approach is certainly feasible though it requires the
introduction of an additional parameter, $\kappa_b$, in the
model. With the purpose of keeping the model as transparent as
possible we have restricted the analysis to the simplest formulation
of the TC. This allows to ascertain how the reduction of
configurational entropy operated by the thickness constraint affects a
chain elastic response. Indeed, the restriction on the allowed
conformational space is so severe that it is able to induce a finite
persistence length, $\xi_p$ in the modelled chains. 

{At present, no exact expression for the dependence of
$\xi_p$ on $\Delta$ and $a$ is available. However, a useful
approximate expression can be obtained by retaining only the local
thickness constraints and neglecting the non-local ones. In this case,
the only configurations available to the TC are those where the cosine
of the angle formed by the {tangent vectors}
of two consecutive chain bonds,
$\vec{t}_0$ and $\vec{t}_1$ is greater than $1 - a^2/2\Delta^2$. The
average value of their scalar product is therefore, $\langle \vec{t}_0
\cdot \vec{t}_1 \rangle = 1-a^2/4\Delta^2$. Subsequent bond 
{tangent vectors},
$\vec{t}_n$ will accumulate increasing deviations from the first
{tangent vector}: $\langle \vec{t}_0 \cdot \vec{t}_n \rangle =
(1-a^2/4\Delta^2)^n$ \cite{dnapack}. One thus obtains that the
persistence length induced by the local thickness constraint is}

\begin{equation}
\xi_p=-\frac{a}{\log\left(1-\frac{a^2}{4\Delta^2}\right)}\ .
\label{eqn:xi}
\end{equation}

\noindent {The previous analysis has strong analogies
with the approach of Premilat and Hermans and of Flory and Conrad
 \cite{premilat1,flory10} which showed the profound influence of
dihedral angles restrictions on the persistence length of poly-Gly and
poly-Ala chains.} The relevance of the thickness constraint for the
modelling of biopolymer elasticity can be illustrated by applying
eq. (\ref{eqn:xi}) to the case of polypeptides and DNA. In the first
case the chain modularity is intuitively suggested by the separation
of consecutive $C_\alpha$ atoms, $a=3.8$ \AA, while the typical
polypeptide thickness obtained from previous studies is $\Delta =
2.5$\AA. This yields for $\xi_p$ the value of 0.55 nm which is in good
agreement with the experimental value of 0.4 nm
 \cite{gaub97,gaub99}. For single- and double-stranded DNA $a$ is
naturally taken as the nucleotide separation ($a_{ss} = 0.55 $nm) or
the base-pair spacing, ($a_{ds} = 0.34 $nm), respectively, while the
associated nominal values for $\Delta$ based on structural
determinations are $\Delta_{ss} = 0.4$ nm and $\Delta_{ds} = 1.25$ nm.
For single-stranded DNA this yields $\xi^{ss}_p \approx 1$ nm, which
is consistent with experimental determinations in solutions rich of
monovalent counterions which screen the phosphate
charges \cite{WLC1}. For double-stranded DNA one obtains
$\xi_p^{ds} = 19 $ nm which captures the correct order of magnitude of
the experimentally-determined one $\xi_p = 50$
nm. 
{These examples, though based on the simplified
local approximation for the persistence length, illustrate well how
the basic elastic properties of various biopolymers can be influenced
by their finite thickness.}

In the present study the complete characterization of the stretching
response of a thick chain was obtained through extensive Monte Carlo
simulations. The discretization length, $a$, was taken as the unit
length in the problem and several values of $\Delta/a$ were
considered, ranging from the minimum allowed value of 0.5
(corresponding to the limit case where the oblate ellipsoids of
Fig. 1 are spheres) to the value of 4.0. This upper limit
appears adequate in the present context since the largest nominal
ratio for $\Delta/a$ among the biopolymers considered here is achieved
for dsDNA for which one has $\Delta/a \approx 3.7$. For each
value of $\Delta/a$ considered, the simulations were carried out on
chains whose length was at least ten times bigger than the persistence
length estimated from eq. (\ref{eqn:xi}). The relative elongation of
the chain was calculated for increasing values of the applied
stretching force (typically about 100 distinct force values were
considered). Starting from an arbitrary initial chain configuration,
the exploration of the available structure space in the fixed-force
ensemble was done by distorting conformations by means of pivot and
crankshaft moves. The standard Metropolis algorithm was used to
accept/reject the newly generated conformations. For each run, after
equilibration, we measured the autocorrelation time and sampled a
sufficient number of independent conformations to achieve a relative
error of, at most, $10^{-3}$ on the average chain elongation. For
moderate or high forces this typically entailed the collection of
10$^4$ independent structures while a tenfold increase of sampling was
required at small forces due to the broad distribution of the
end-to-end separation along the force direction.

{From the simulations several different stretching
regimes could be identified in the elastic response of a thick
chain. These are best discussed in connection with analogous regimes
found in the FJC and WLC. For the former model, the functional
dependence of the relative elongation, $x$, on the applied force, $f$,
is given by}
 
\begin{equation}
{x = {\rm coth}(f\,b/k_B T) - {k_B T \over f b} }
\label{eqn:fjc}
\end{equation}

\noindent {where $b$ is the Kuhn length, and $k_B\,T$
is the thermal energy. The analogous characteristic relationship for
the WLC is, instead, traditionally approximated as:}

\begin{equation}
 f(x) = \frac{k_BT}{\xi_p}\left[\frac{1}{4(1-x)^2}-\frac{1}{4}+
{x}\right]
\label{eqn:wlc}
\end{equation}

\noindent { where $\xi_p$ is the persistence length. It
is apparent from both expressions that, at low force, the relative
elongation, $x$, depends linearly on the applied force, $f$. This
result holds also for the thick chain model. However, due to the
self-avoidance of the TC, the Hookean relationship between $f$ and $x$
disappears in the limit of long polymer chains in favour of the Pincus
regime, $f\sim x^{1/(1/\nu-1)}$, $\nu \sim 3/5$ being the critical
exponent for self-avoiding polymers in three
dimensions \cite{flory,pincus,lam}.} For intermediate forces the
Pincus behaviour is found to be followed by a second regime
characterised by the same scaling relation found in the WLC at high
forces, $f\sim (1-x)^{-2}$. As shown in Fig. 2, at still higher forces
the same scaling relation found in the FJC is observed, $f\sim
(1-x)^{-1}$. Physically, the first two regimes are determined by
self-avoidance and chain stiffness or persistence length, while the
last regime, found in the Kratky-Porod
model, has recently been ascribed to the discrete nature of the
chain \cite{netz,nelson,angelo,lamura}.

Since the purpose of this study is to apply the thick-chain model to
contexts where experimental data are available we have analysed the
numerical data with the purpose of extracting a single analytical
expression which captures the observed functional dependence of $f$ on
$\Delta$ and $a$. For any value of $a$ and $\Delta$ the sought
expression should reproduce the succession of the three regimes
discussed above. Among several trial formulae we found, {\em a
posteriori}, that the best interpolation was provided
by the following expression,

\begin{equation}
f(x)=\frac{k_BT}{a (1 -x)}
\tanh{\left(\frac{k_1x^{3/2}+k_2x^2+k_3x^3}{1-x}\right)}
\label{analytic_approximation}
\end{equation}
where the parametric dependence on $\Delta$ and $a$ is carried by the
following expressions for the $k$'s,

\begin{eqnarray}
k_1^{-1} & = & -0.28394 + 0.76441\, {\Delta}/{a} + 0.31858\, {\Delta^2}/{a^2},\\
k_2^{-1} & = & +0.15989 - 0.50503\, {\Delta}/{a} - 0.20636\, {\Delta^2}/{a^2},\\
k_3^{-1} & = & -0.34984 + 1.2333\, {\Delta}/{a} +
0.58697\, {\Delta^2}/{a^2}.
\end{eqnarray}

Within the explored ranges of $\Delta$ and $a$ the relative extension
obtained from eq. (\ref{analytic_approximation}) differs on average by
1\% (and at most by 5\%) from the true values at any value of the
applied force. Through the analysis of the decay of tangent-tangent
correlations in the configurations sampled in the absence of a
stretching force we also obtained the persistence length as a function
of $\Delta$ and $a$. For this quantity we found that expression of
eq. (\ref{eqn:xi}) already provides a very good approximation for the
observed persistence length and found no need to correct it with other
interpolation formulae.

{In the discrete chain model considered here, the
spacing of consecutive beads is constant along the chain. The
application of arbitrary large forces cannot, therefore, stretch a
thick chain beyond its nominal contour length. This is true also for
the FJC and WLC models described by equations (\ref{eqn:fjc}) and
(\ref{eqn:wlc}), since $x$ cannot exceed 1 for any value of $f$. The
property of inextensibility, though very natural and appealing, can
make the models inadequate to capture the phenomenology of stretching
at forces high enough to cause isomerization or structural
transitions. Experiments on dsDNA, for example, have shown that forces
greater than 10 pN can induce significant overstretching in the
molecule \cite{block1}. For polysaccharides such as dextran, a
chair-to-boat transition occurs at forces around 700 pN, leaving a
distinctive ``bump'' in the force versus elongation curves. The
detailed description of how each type of biomolecule distorts at high
stretching forces requires the use of {\em ad hoc} models and atomistic
simulations thus going beyond the reach of simple phenomenological
models \cite{marsza4}. However, the simple inextensible models
mentioned above can be extended to describe, in a very satisfactory
way, also the over-stretching regime. This is accomplished by
introducing an effective stretching modulus, $K$, controlling the
elongation of the chain beyond its natural contour
length \cite{odijk1}. From a practical point of view the relative
extension, $x$, of eqns. (\ref{eqn:fjc}) and (\ref{eqn:wlc}) is
replaced by $x\,(1 - f/K)$. The same scheme could be adopted also with
the TC model too. In the interest of limiting the number of parameters
in the theory, the focus of the present work is limited to the
inextensible model which has, so far, been the reference case for 
thick chain models.}

\section{Results and Discussion}

We have applied the thick-chain results of eq.
(\ref{analytic_approximation}) to interpret the experimental
stretching curves of three types of biomolecules: the PEVK domain of
titin, cellulose and double-stranded DNA. 
{Due to the
inextensible nature of TC model, it will be applied to fit data
collected at forces below the known overstretching thresholds
mentioned below.} Our goal is twofold: on one hand we wish to verify
the viability of the thick-chain model in contexts where other
entropic models are routinely used. Secondly, we aim at extracting the
relevant structural information, 
{(effective thickness
and granularity)}, of the polymer from the experimental stretching
curves.

\noindent For both purposes a first useful benchmark is constituted by
the PEVK-domain of titin. The experimental data which we shall
consider pertain to the atomic force measurements of ref.
 \cite{linke} that were performed on heteropolymers comprising cardiac
PEVK segments (each about 190 residues long) in tandem with I27
domains. Some evidence has been previously presented in favour of the
lack of structural organization in PEVK domain. It therefore appears
appropriate to model the PEVK elastic response as arising from
entropic effects, as in a random coil (to which the PEVK domain has
indeed been assimilated). The use of the WLC to fit the low-force
data on the hetero-polyprotein returned a persistence length spanning
the range 0.5 - 2.5 nm, the average being 0.9 nm \cite{linke}.

Fig. 3 reproduces the stretching measurements on the polyprotein
fitted by the TC model (assuming a constant error on the experimental
force). To avoid the introduction of the spurious effects arising from
the I27 mechanical unfolding (at 200 pN), in the fit of Fig. 3 and in
all other sets of measurements made available to us we considered
forces only up to 100 pN. For the specific example portrayed in Fig. 3
the contour length associated to the low-force part of the first peak
resulted equal to 147.6 $\pm$ 0.3 nm, fully consistent with the one
determined from the WLC, 148.1 nm $\pm$ 0.1. The inferred structural
parameters were, instead, $a= 0.45 \pm 0.02$ nm, $\Delta = 0.34 \pm
0.01$ nm which, from eq. (\ref{eqn:xi}) yield a persistence length of
0.77 $\pm$ 0.03 nm. Again, this value is consistent with the one
obtained from the WLC on this particular sample, $\xi_p^{WLC} = 0.76$
$\pm$ 0.02 nm. The advantage of the present approach is, however, that
some insight is provided about the main structural parameters of the
polymer. The average values of $\Delta$ and $a$ obtained from the fits
of several experimental instances of (I27-PEVK) repeats gave $\Delta =
2.63 \pm 0.36$ {\AA} and $a = 3.58 \pm 0.35$ \AA. Both these
parameters appear to capture well the structural features of
polypeptides. In fact, the effective separation of the modular grains
is of the same order of the natural coarse-graining provided by the
separation of consecutive amino acids, $a=3.8$\AA\, though this value
is just indicative in the present context where it is not known what
fraction of the numerous prolines are in cis/trans
conformations. Furthermore, the effective thickness of $\Delta = 2.63
\pm 0.36$ {\AA} is in excellent agreement with the one obtained by
direct processing of several independent protein structures, $\Delta
\sim 2.7$ \AA \cite{secstr}. Finally, we mention that the statistical
significance ($\chi^2$) of the TC fit is almost exactly the same as
for the WLC. In this context the FJC is the worst performer since the
associated $\chi^2$ exceeds the TC or WLC one by a factor of two.

{Another relevant class of biopolymers for which
stretching measurements are available is constituted by
polysaccharides, in particular cellulose, amylose and dextran. We
shall focus on cellulose since its stretching curve does not display
the presence of the ``bumps'' which in amylose and dextran denote the
onset of structural transitions which can be captured only by
atomistic models. On the contrary, cellulose, which is composed of
sugar rings each spanning a distance of about 1 nm, does not
display dramatic structural changes upon stretching. For
these reasons it is regarded as a standard example of entropic
elasticity.} It should be mentioned that, within traditional elastic
models, the best fit of experimental data for forces up to 300 pN, is
provided not by the WLC but by the FJC. In fact, under the working
assumption of a constant error on the measured force (but also using
an error proportional to the force) the FJC fit provides a $\chi^2$
smaller than the WLC by more than 50\%. The lengths
of the statistical segments obtained from the FJC fit is $l_K = 1.03
\pm 0.02$ nm which is very close to the nominal lengths of the sugar
units comprising cellulose. Consistently, the application of the TC
model (Fig. 4) yields $a=1.02 \pm 0.01$ nm and $\Delta = 0.54 \pm
0.02$ nm which are both in agreement with the atomic structural
parameters of cellulose.

Finally, we consider the case of dsDNA stretching. The extensive
experimental data set used here was obtained by optical tweezers
measurements on DNA in a PTC buffer \cite{wang}. The type and
concentration of the counterions in solution are known to affect
strongly the strength (and even the sign) of the self-interaction of
DNA molecules through the screening of the DNA
phosphates \cite{cozzarelli,odjik,podg00}. In a seminal study
Cozzarelli {\em et al.} have described the effects of the
self-repulsion of DNA in a solution containing monovalent counterions
(Na$^+$) in terms of an apparent DNA diameter \cite{cozzarelli}. In
their study this effective diameter was recovered from the comparison
of the knotting probabilities found experimentally for circular DNA of
known length with those predicted numerically for a circular
arrangements of a succession of cylinders. It was found that the
effective diameter approached the bare-DNA one (2.5 nm) for high
concentrations of counter-ions, while for low concentrations the poor
electrostatic screening of the phosphates yielded much greater
diameters, e.g. 15 nm in the presence of 10 mM Na$^+$. Our approach
provides an alternative framework for the extraction of the effective
DNA diameter which, furthermore, is inferred not from the analysis of
structural properties in an equilibrium ensemble (as for the knotting
probabilities) but from the elastic response of single molecules.

The good performance of the TC model (the fit is shown in Fig. 5) is
highlighted by the very good value for $\chi^2$ which optimizes the
fit. Even considering measurements up to forces of 25 pN (more than
650 data points) the $\chi^2$ takes on the excellent value of
1.2. This constitutes a significant improvement among the class of
inextensible chain models which are usually employed only up to forces
of 10 pN. In fact, the performance of the inextensible WLC, both in
its original and improved form, up to forces of 25 pN yields a value
for $\chi^2$ greater than 2. Within the WLC scheme only through the
addition of enthalpic effects which mimic the chain extensibility it
is possible to improve the quality of the fit reaching the value
$\chi^2=1.1$. Incidentally, we mention that the FJC yields a poorer
fit than the WLC or TC.

The fit through the TC model yields a contour length of 1.335 $\pm$
0.001 $\mu$m, very close to the one recovered from
the original WLC model 1.327 $\pm$ 0.001 $\mu$m (using forces up to 10
pN) or to the extensible version using forces up to 25 pN. The
structural parameters extracted in this context are $a =$ 3.27 $\pm$
0.57 nm and $\Delta$ = 5.76 $\pm$ 0.58 nm. They correspond to a
persistence length of 38.99 $\pm$ 0.86 nm, again consistently with
e.g. the one obtained from the improved WLC, $\xi_p = 37.8$ $\pm$ 0.5
nm. Consistently with the findings of ref.  \cite{block1} the use of
the extensible WLC model leads to a slightly larger persistence
length, $\xi_p=40.6$nm. 
{Arguably, the smaller $\chi^2$
 values obtained with the TC model may reflect the presence of an
 additional parameter with respect to the worm-like or freely-jointed
 chains. However, the increase of parameter space does not
 automatically lead to substantial improvements of the fits. If we
 were to refine the WLC model with its discretised
 version \cite{angelo} whose parameters would then be the contour
 length, persistence length, and granularity, we would not improve
 the value of $\chi^2$ obtained with the ordinary WLC model. The
 improvement obtained with the TC model can therefore be ascribed to
 the ability to capture in a better way the phenomenology of polymer
 stretching (in the same way that extensible models augment the
 range of applicability of inextensible ones).}

The consistency of $L_c$ and $\xi_p$ with the data obtained from the
WLC shows that the familiar elastic parameters of the chain are
faithfully recovered by the thick chain model too. The additional
insight obtained here is the indication of the effective DNA
granularity and thickness. In fact, the value $a= 3.27$ nm corresponds
to about nine times the nominal base-pair spacing in dsDNA while the
effective thickness, which corresponds to a diameter $2\Delta \approx
11.5 \pm 1$ nm is about five times the nominal hydration radius of
dsDNA. A direct quantitative comparison of this value with the
effective diameters found in ref. \cite{cozzarelli} cannot be done
with the present data set given the different type of solution
employed. This comparison is, however, possible using a set of DNA
stretching measurements done in the presence of sodium ions
 \cite{busta,thiru}. Also in these cases the TC model provides a better
$\chi^2$ than the FJC or WLC. It is found that in 10 mM Na$^+$
the effective diameter is $2 \Delta = 23 \pm 3$ nm which exceeds by 35
\% the estimate obtained by Cozzarelli {\em et al.} from the
probability of occurrence of trefoil knotted DNA configurations. Given
the diversity of the two approaches it is pleasing that the two types
of effective diameters are in fair quantitative agreement.
Consistently with the intuitive expectations and the findings of
ref.~\cite{cozzarelli} it is found that at lower ionic strengths the
effective diameter increases so that e.g. $2\Delta = 51 \pm 7$ nm at
1.0 mM. 
{Of all the cases considered here, the aspect
that is less satisfactory is the order-of-magnitude difference between
the effective spacing of the beads and the base pair
separation. Although this consideration is mostly based on intuition,
the good agreement of the effective bead spacings and nominal length
of the ``moduli'' found for PEVK and polysaccharides suggests that a
better consistency should be obtained for DNA too. This might be
accomplished by considering, in addition, to the thickness
constraints, other important factors limiting the local freedom of the
chain, such as introducing a bending or torsional rigidity term
(see ref.~\cite{jpa_05} for a theoretical framework within which
to consider this effects). Such
extension of the TC model will be the object of future
investigations.}


\section{Conclusions}

In summary, we have studied the feasibility of extracting the
effective thickness and granularity of a biopolymer by fitting its
stretching response to that of a discrete thick polymer. The strategy
that we followed relied on the preliminary numerical characterization
of the elastic response of a thick chain through several Monte-Carlo
simulations. From the numerical results we extracted a simple
functional form for the force versus elongation curve as a function of
effective thickness, chain granularity, and contour length. The
resulting expression was used to analyse data coming from stretching
experiments on the PEVK domain of proteins, cellulose and double
stranded DNA in presence of different concentrations of monovalent
counter-ions. For all these biopolymers, we thus recovered, besides
the persistence length (which is consistent with that obtained from
standard entropic models) quantitative indications of their thickness
and effective monomer size. These structural parameters turned out to
be compatible with the nominal ones for the PEVK-domain and cellulose
or, in the case of DNA, with those previously obtained through a
careful analysis of the proportion of knotted DNA molecules.

The results clarify that the use of suitable models allows to
extract from the experimental stretching curve not only some overall
elastic phenomenological parameters but also fundamental
structural properties of the polymer itself. This issue is not
only a challenge in itself but has important experimental
ramifications in the characterization of the polyelectrolytes
self-interaction. The simple approach proposed here in terms of the
thick chain model constitutes a novel effective scheme for
characterizing the effective size of polyelectrolytes in various
ionic solutions, starting from measurements of their elastic
response.

{\it Acknowledgments:} We are indebted to S. Block, W. Linke and
P. Marszalek for having provided us with the experimental data that were
used in this study. We are grateful to E. Lattman, W. Linke and
A. Pastore for fruitful discussions. This work was supported by INFM.

\begin{figure}
\includegraphics[width=4.0in]{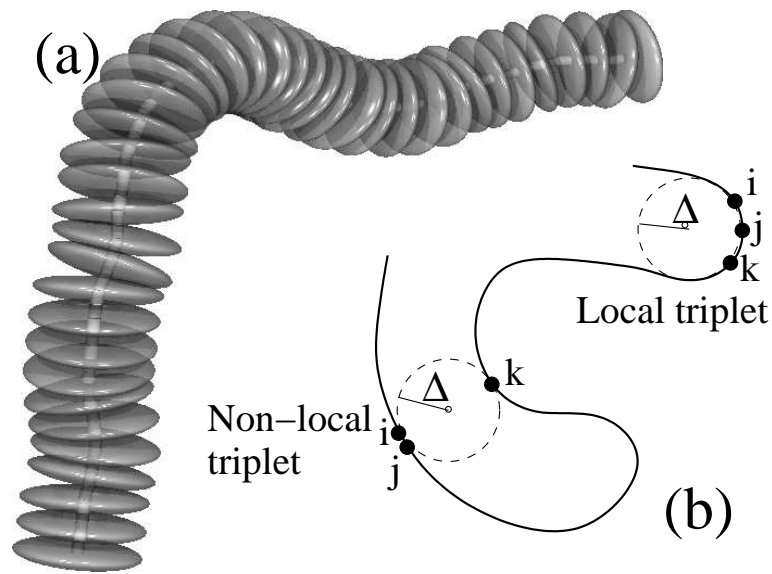}
\caption{(a) Pictorial sketch of a discrete thick chain with thickness $\Delta$
and granularity $a$ (in the figure $\Delta=4a$). The finite
thickness introduces steric constraints that forbid configurations
where the chain self-intersects. (b) These constraints, which for,
e.g. the chain of ellipsoids, may be implemented only with heavy
computational expenditure, are more conveniently treated within the
three-body prescription of the thick-chain model. Within this
approach the centerline of a viable configuration is such that the
radii, $r_{ijk}$ of the circles going through any triplet of points,
$i,\ j,\ k$, on the curve  are not smaller than $\Delta$}
\label{fig:1}
\end{figure}

\begin{figure}
(a)\includegraphics[width=2.5in]{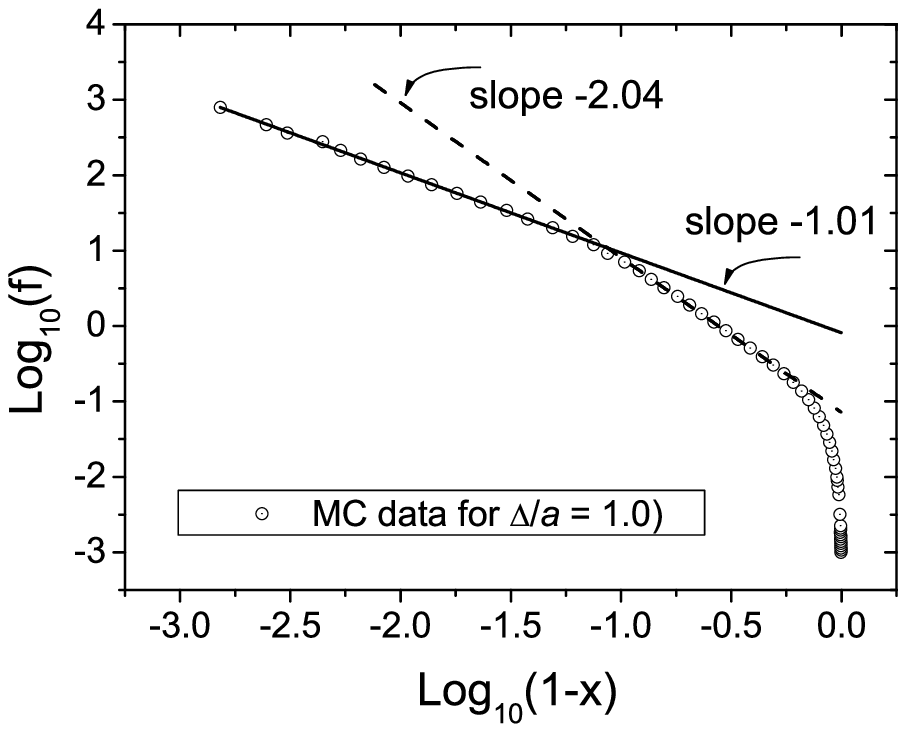}
(b)\includegraphics[width=2.5in]{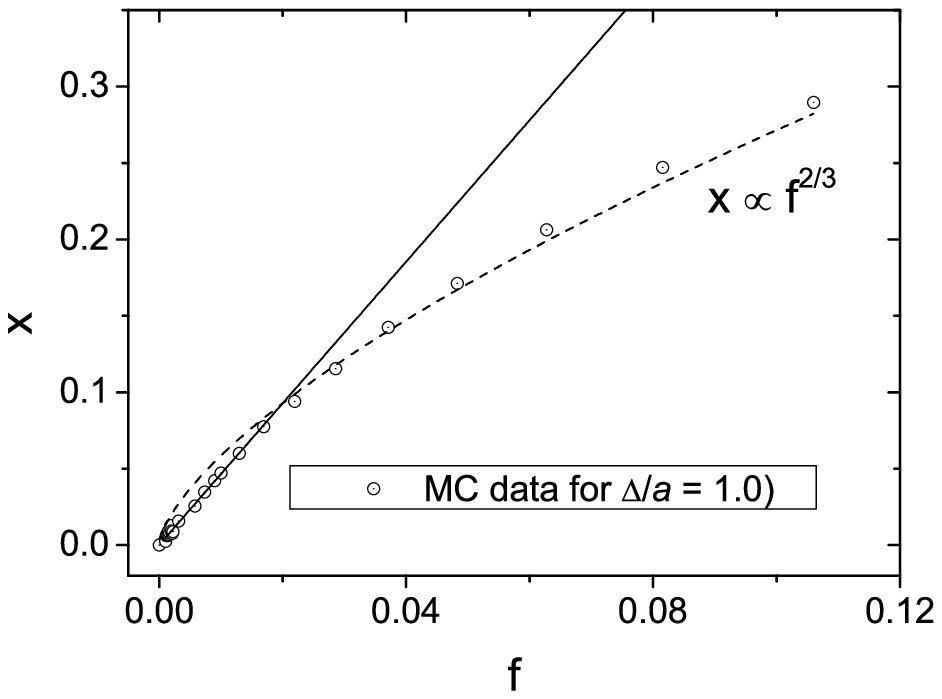}
\vskip 0.8 cm
\centerline{Figure 2}
\caption{(a) Elongation versus reduced force for a chain of
thickness $\Delta/a=1$. Data points are presented in the
$\left(\log(1-x),\log(f)\right)$ plane to highlight the WLC- and
FJC-like regimes found at moderate and high forces. (b) Illustration
of the low-force crossover from the Hookean regime, $x \propto f$, to
the Pincus one, $x \propto f^{2/3}$ for a chain of thickness
$\Delta/a=1$ and $N=1200$ beads.}
\label{fig:3}
\end{figure}

\begin{figure}
\includegraphics[width=4.0in]{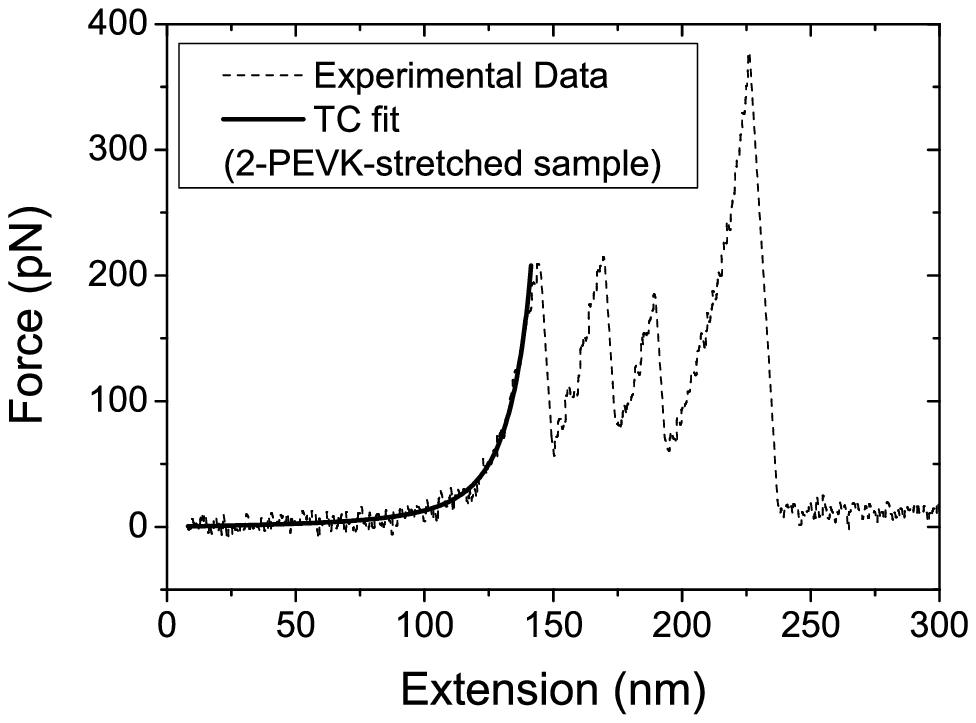}
\vskip 0.8 cm
\centerline{Figure 3}
\caption{Typical fit of the PEVK stretching data through the
thick-chain model. The particular fit yields a contour length $L_c
= 147.6 \pm 0.3$ nm, a thickness $\Delta = 0.34 \pm 0.01$ nm and a
granularity $a = 0.45 \pm 0.02$ nm. For this set the WLC provides
a fit with a $\chi^2$ analogous to that of the TC and yields $L_c
= 148.1 \pm 0.1$ nm and $\xi_p = 0.76 \pm 0.02$ nm.}
\label{fig:PEVK}
\end{figure}

\begin{figure}
\includegraphics[width=4.0in]{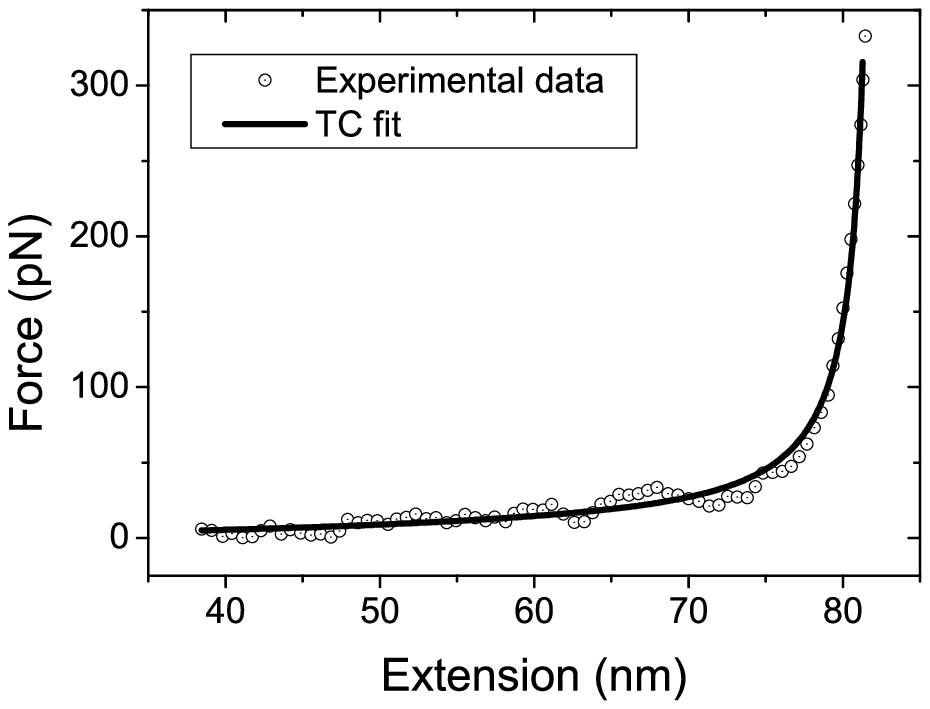}
\vskip 0.8 cm
\centerline{Figure 4}
\caption{Fit of the cellulose stretching data through the TC
model. The fit was carried out using forces up to 300 pN, and
yielded a contour length $L_c = 82.35 \pm 0.02$ nm, a thickness
$\Delta=0.54 \pm 0.02$ nm, and a granularity $a = 1.02 \pm 0.01$ nm.}
\label{fig:cellulose}
\end{figure}

\begin{figure}
\includegraphics[width=4.0in]{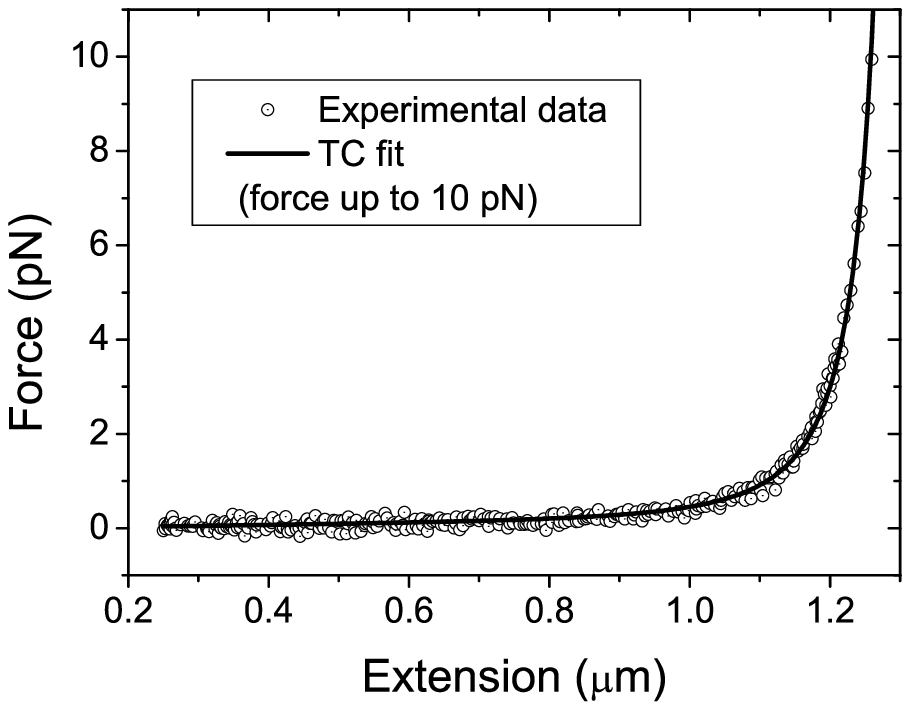}
\vskip 0.8 cm
\centerline{Figure 5}
\caption{Fit of dsDNA stretching measurements through the TC
model. Forces up to 10 pN were used.}
\label{fig:dsDNA}
\end{figure}
\end{document}